\documentclass[sigconf, nonacm]{acmart}
\AtBeginDocument{%
  \providecommand\BibTeX{{%
    \normalfont B\kern-0.5em{\scshape i\kern-0.25em b}\kern-0.8em\TeX}}}

\copyrightyear{2021} 
\acmYear{2021} 
\setcopyright{acmlicensed}\acmConference[DIS '21]{Designing Interactive Systems Conference 2021}{June 28-July 2, 2021}{Virtual Event, USA}
\acmBooktitle{Designing Interactive Systems Conference 2021 (DIS '21), June 28-July 2, 2021, Virtual Event, USA}
\acmPrice{15.00}
\acmDOI{10.1145/3461778.3462096}
\acmISBN{978-1-4503-8476-6/21/06}

\usepackage{relsize}
\usepackage{xspace}
\graphicspath{{Images/}}

\newcommand{\etal}{\textit{et al.}\xspace}
\usepackage[squaren,thinqspace,textstyle]{SIunits}

\begin{document}

\title{SelectVisAR: Selective Visualisation of Virtual Environments in Augmented Reality} 


\author{Robbe Cools}
\authornote{Both authors contributed equally to this research.}
\affiliation{%
  \institution{KU Leuven}
  \department{Department of Computer Science}
  \city{Leuven}
  \country{Belgium}
}
\email{robbe.cools@kuleuven.be}

\author{Jihae Han}
\authornotemark[1]
\affiliation{%
  \institution{KU Leuven}
  \department{Department of Computer Science}
  \city{Leuven}
  \country{Belgium}
}
\email{jihae.han@kuleuven.be}

\author{Adalberto L. Simeone}
\affiliation{%
  \institution{KU Leuven}
  \department{Department of Computer Science}
  \city{Leuven}
  \country{Belgium}
}
\email{adalberto.simeone@kuleuven.be}


\begin{abstract}

When establishing a visual connection between a virtual reality user and an augmented reality user, it is important to consider whether the augmented reality user faces a surplus of information. Augmented reality, compared to virtual reality, involves two – not one – planes of information: the physical and the virtual. We propose SelectVisAR, a selective visualisation system of virtual environments in augmented reality. Our system enables an augmented reality spectator to perceive a co-located virtual reality user in the context of four distinct visualisation conditions: \textit{Interactive}, \textit{Proximity}, \textit{Everything}, and \textit{Dollhouse}. We explore an additional two conditions, \textit{Context} and \textit{Spotlight}, in a follow-up study. Our design uses a human-centric approach to information filtering, selectively visualising only parts of the virtual environment related to the interactive possibilities of a virtual reality user. The research investigates how selective visualisations can be helpful or trivial for the augmented reality user when observing a virtual reality user. 

\end{abstract}




\begin{teaserfigure}
  \includegraphics[width=\textwidth]{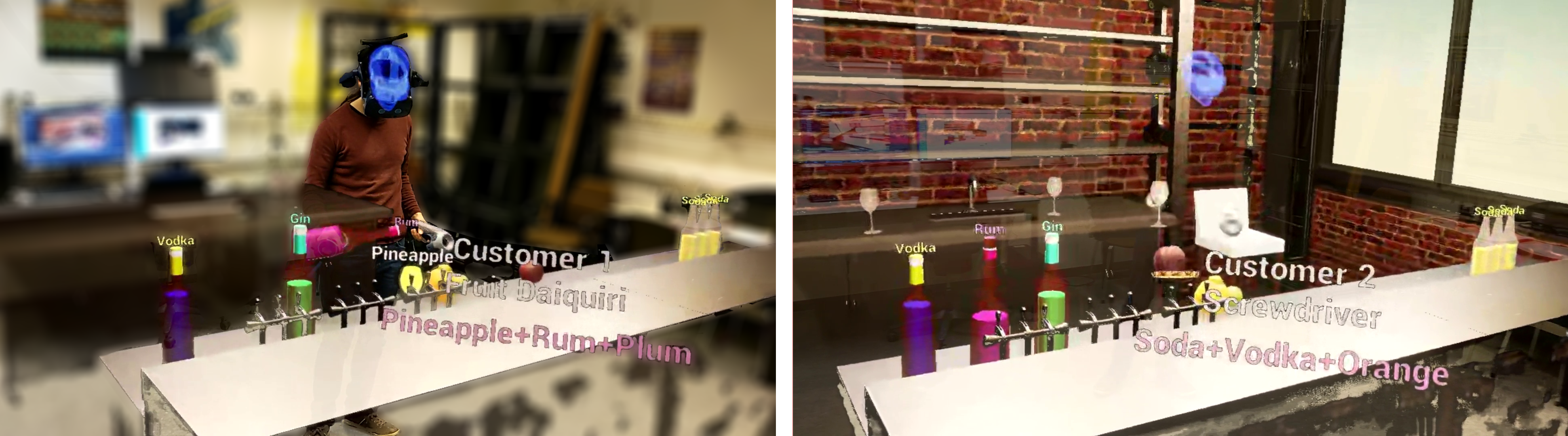}
  \caption{\textit{Context} static selection technique (left) and \textit{Everything} technique (right), AR User Perspective}
  \Description{}
  \label{fig:teaser}
\end{teaserfigure}

\maketitle

\section{Introduction}
Virtual Reality (VR) and see-through Augmented Reality (AR) devices are becoming increasingly affordable. VR enables users to immerse themselves in a Virtual Environment (VE). A see-through AR device can overlay virtual content on top of the physical environment. VR and AR users can interact with each other through Collaborative VEs ~\cite{Xia2018SpaceTime:Reality}, and the interaction between a VR and AR user is considered a type of "Cross-Reality Interaction". 

Cross-Reality (CR) is a field of research that looks at how users of different realities can interact with each other. These realities can be described through Milgram's Reality-Virtuality continuum~\cite{Milgram1995AugmentedContinuum,Skarbez2021RevisitingContinuum}, which ranges from the real world to the  virtual world and the spectrum of hybrid realities in between. This paper will focus on  scenarios that involve interactions between two different points in this continuum: AR and VR. 

This type of scenario can be beneficial when the roles of the AR and VR user in the collaboration are asymmetrical. It is important to note the differences in how users perceive their VEs: VR users benefit from more immersion and AR users benefit from more nonverbal cues. In a CR context, nonverbal cues refer to the advantage AR users have over VR users when communicating with an external user — For instance, both VR and AR users can talk to an external user, but only AR users can see the physical body and gestures of an external user in real life. The VR user cannot see the external user, and at most can only perceive the external user's virtual avatar. As such, AR retains most nonverbal cues lost to VR users. In contrast, while VR users retain high immersion, AR users will be less immersed in the VE due to their vision of the physical environment. 

For some users VR might be more desirable, such as for a training simulation where the user needs to have a sense of being at the place of the training. Other users can benefit from AR to enable them to see nonverbal cues of other co-located users. In this study, we aim to develop a CR scenario that exploits both the immersive benefits of VR and the nonverbal communication features of AR. We question how an AR user can spectate and interact with the VR user in their VE~\cite{Han2020TheObjects}. We propose a selective visualisation system that enables AR users to only see select virtual elements of the VE whilst VR users see the entire VE to maintain their immersion. 

We investigated two design factors in terms of visualising a VE: level of information and scale. We designed a framework where we presented the VE to the AR user at a 1:5 dollhouse-scale and at 1:1 room-scale with three levels of information: no selection, a dynamic selection following the VR user, and a predetermined static selection. This study was then repeated with two improved dynamic and static selection techniques implementing feedback from the main study.

In our studies we found that participants felt they had a better overview of the VE at the small 1:5 dollhouse-scale; however, this had the drawback that it was more difficult to see smaller movements of the VR user. We found that our dynamic selection methods were preferred by fewer participants than the static selections. No significant differences were found in participant competence in recognising events in the VE.

\section{Related Work}

CR collaboration refers to users on different points on the Reality-Virtuality continuum~\cite{Milgram1995AugmentedContinuum,Skarbez2021RevisitingContinuum} working together. In order to support collaboration, users need to be aware of each other's ``realities'' and be able to interact with other users and their reality.

\subsection{Tablet and screen-based VE Visualisation}
Different technologies can be used to support CR visualisation and interaction. TransceiVR~\cite{Kumaravel2020TransceiVR:Collaborators} enables a non-immersed tablet user to view the VE from the perspective of the immersed user by freezing the frame and making annotations that are communicated back into the VE. The real environment is disconnected from the VE, as the external user sees it from the perspective of the immersed user. FaceDisplay~\cite{Gugenheimer2018FaceDisplay:Reality} mounted screens on the Head-Mounted Display (HMD), through which the external user can view the VE. This presents the VE from the perspective of the external user, however only when they are looking directly at the VR user's HMD.

Silhouette Games~\cite{Krekhov2020SilhouetteVR} presents an approach with a screen behind a one-way mirror. The screen displays a simulated reflection of the VE calculated based on the position of the non-immersed user. The non-immersed user can then view the VE and the physical reflection of the VR user simultaneously. Seeing both the VR user and their reflection caused some confusion in participants. Our AR-based approach does not rely on a reflection of the VR user, but visualises the VE directly around them. This does require the external user to wear an AR HMD which is more invasive than the approach presented in Silhouette Games.

\subsection{Projection-based VE Visualisation}
 Wang \etal~\cite{Wang2020HMDInteraction} mounted a projector on the HMD, which allows visualisation of the VE on the floor around the immersed user. The VR user had control over the content that was shown in the projection. In this paper we investigate techniques that visualise the area of the VE around the VR user, changing the visualisation as the VR user moves.
 
 ShareVR~\cite{Gugenheimer2017ShareVR:Users} combined a static floor projection, covering the entire space available to the VR user, with a handheld screen to enable interaction between an immersed user and an external user. ReverseCAVE~\cite{Ishii2019LetExperience} also used a projection-based approach, where the VE was not projected on the floor but on four translucent screens around the VR user that external users can then spectate from the outside. In our work we also investigated static visualisations covering the entire available space, however we used see-through AR instead of a projection.
 
\subsection{Miniatures}
Pham \etal~\cite{Pham2018ScaleDesign} investigated the effect of the scale of AR visualisation on gestures, investigating models at `in-air' scale, tabletop scale and room scale. They found that these different scales elicited different gestures from users. We will investigate the effect of the scale of the visualisation on an AR user spectating a VR user, inspired by Dollhouse VR~\cite{Ibayashi2015DollhouseTechnology} and World In Miniature~\cite{Pausch1995NavigationMiniatures}. We find further investigation on scale relevant as neither Dollhouse VR \cite{Ibayashi2015DollhouseTechnology} nor its follow-up study \cite{Sugiura2018AnDesign} specifies or justifies the use of a specific scale when implementing the `dollhouse', only detailing a relative size difference between visualisations. World In Miniature provides more but still relatively abstract detail regarding its implementation of scale, remarking that a World In Miniature may be `hand-held' but not specifying a scalar value \cite{Pausch1995NavigationMiniatures}.

\subsection{AR-based VE Visualisation}

Grandi \etal~\cite{Grandi2019CharacterizingRealities} investigated collaboration between VR and tablet AR users. AR and VR users were co-located and collaborated on solving a docking task. The AR user saw the virtual objects with the same spatial orientation as the VR user. The shared virtual elements were limited to tabletop-size objects, whereas we investigated how to visualise the VR user in context of their VE to the AR user on room-scale.

 ObserVAR~\cite{Thanyadit2019ObserVAR:Reality} explores the use of see-through AR for a teacher to visualise their students, who are immersed in a VE using 3-DOF VR devices. Three visualisations were tested: \textit{First Person View}, \textit{World in Miniature} and \textit{World Scale}. Participants found \textit{World Scale} easier to use than \textit{World in Miniature}, though scale was not the only factor because \textit{World in Miniature} showed a separate miniature per VR user. The ObserVAR user study had multiple remote VR users, while our study had a single co-located VR user and focused on visualising them in context of the VE.
 
\subsection{Selection within VEs}
Slice of Light~\cite{Wang2020SliceEnvironment} presents a method for a VR user to see and move between VEs of other VR users. The other users' VEs are visualised as slices around the user in that VE, the external user can then enter that VE be stepping towards it. The purpose of presenting the VEs as slices is so that multiple users in their VE can be shown at once. We investigated if filtering what is shown of the VE can improve AR users' understanding of the VR user's actions. To do this we tested both static and dynamic selections of virtual content. A summary positioning our work to the related work can be seen in \autoref{tab:summary}. The table characterises the visualisation of the VE according to the Point of View (PoV) from one or a combination of VR, AR, and Physical Reality (PR) users. 

\begin{table}[ht]
\begin{tabular}{ |c|c|c| } 
 \hline
 \textbf{Interaction} & \textbf{VE Visualisation} & \textbf{Name} \\ 
 \hline
  PR-VR & VR user's PoV (tablet) & \textit{TransceiVR} 
  \cite{Kumaravel2020TransceiVR:Collaborators}\\ \hline
  PR-VR & VR user's PoV (tablet) & \textit{FaceDisplay} \cite{Gugenheimer2018FaceDisplay:Reality} \\ \hline
  PR-VR & top-down view of scaled& \textit{Dollhouse VR} \cite{Ibayashi2015DollhouseTechnology} \\ 
  &  `dollhouse' VE (tablet) & \\ \hline
  PR-VR & own PoV (tablet) +  & \textit{ShareVR} \cite{Gugenheimer2017ShareVR:Users}\\ 
  & top-down view (projection) &\\ \hline
  PR-VR & VR user's PoV (projection) & \textit{HMD Light}  \cite{Wang2020HMDInteraction} \\ \hline
  PR-VR & 4 perspectives (CAVE) & \textit{ReverseCAVE} \cite{Ishii2019LetExperience}\\ \hline
  AR-VR & AR/VR PoV & CR collab. 
  \cite{Grandi2019CharacterizingRealities} \\ \hline
  AR-VR & `arrow' for VR user's gaze, & \textit{ObserVAR} \cite{Thanyadit2019ObserVAR:Reality}\\\
   & AR/VR PoV, scaled VEs & \\\hline
  AR-VR & selective `filtering' of VE & \textit{*SelectVisAR} \\
   & + scaling of VE & \\ \hline
  VR-VR & selective `slices' of VE & \textit{Slice of Light}~\cite{Wang2020SliceEnvironment}\\ \hline
  VR & world-in-miniature of VE & miniature~\cite{Pausch1995NavigationMiniatures}\\\hline 
  AR & scaling/sizing VEs & holograms~\cite{Pham2018ScaleDesign} \\
 \hline

\end{tabular}

\textit{*SelectVisAR = our own study, in context of related work}

\caption{Table summarising related works.}
\label{tab:summary}
\end{table}

\section{Visualisation and Implementation}
The selection methods we designed aim to visually emphasise the actions of the VR user to an AR spectator by selectively filtering relevant parts of a VE – specifically, the visual artefacts the VR user is interested in or interacting with. This VE is asymmetrically filtered only for the AR user; the VR user would see a fully visualised VE to maintain user immersion. We hypothesise that it is possible to remove a part of the VE without hurting task performance of how an AR user perceives the actions of a VR user. 

\subsection{Pilot Study}
We conducted a pilot study with three HCI experts and a prototype $\unit{5}{\meter}\times\unit{5}{\meter}$ virtual room as the VE. The HCI experts had varying degrees of experience with CR technologies: one expert, one with previous experience, and one without any experience. We conducted a \textit{Think-Aloud Protocol}, with one researcher taking notes and the other assuming the role of the VR user, to prototype our framework and reduce the number of techniques being tested for the main study. We investigated six different visualisation techniques based around the interactive range and possibilities of the VR user, categorised as either static or dynamic visualisations:

Static visualisations select parts of the VE to visualise at all times of the simulation. \textbf{(1)} \textit{Everything} visualises the entire VE to the AR user. This is the control condition in which AR users see the VE in the same way as the VR user. No changes are made to augment or filter information in the visualisation of the VE. \textbf{(2)} \textit{Interactive} is a predetermined selection of interactive objects in the VE. This is inspired by literature that suggests only relevant information should be visualised to prevent overloading the user with irrelevant information \cite{Julier2002InformationReality,Grubert2017TowardsReality}. As `relevant information' is an abstract term, we attempt to draw thresholds in information filtering using \textit{Interactive}. Lastly, \textbf{(3)} \textit{Dollhouse} visualises a smaller, scaled model of the VE. Directly based on Ibayashi \etal~Dollhouse VR \cite{Ibayashi2015DollhouseTechnology}, this visualisation is grounded in previous literature \cite{Pham2018ScaleDesign,Pausch1995NavigationMiniatures} that argue that scaled visualisations of VEs enable more efficient navigation of a VE. However, instead of a 2D top-down view of the VE as investigated in \textit{Dollhouse VR} \cite{Ibayashi2015DollhouseTechnology}, we investigate a 3D scaled model of a VE in AR. 

Dynamic visualisations select different parts of the VE to visualise, depending on where the VR user is located or what the VR user is doing. \textbf{(4)} \textit{Head-Direction} only visualises the part of the VE that the user is facing towards. \textbf{(5)} \textit{Proximity} visualises a radial area of VE nearby the VR user. The conditions \textit{Head-Direction} and \textit{Proximity} are inspired by \textit{Slice of Light} \cite{Wang2020SliceEnvironment}, a visualisation which shows only part of the VE to the guest VR user and dynamically changes depending on the user's location or actions. We based these two conditions on common tracking methods for VR, head-direction tracking for \textit{Head-Direction} and position-tracking for \textit{Proximity}. Lastly,  \textbf{(6)} \textit{Dynamic-Interaction} visualises the virtual objects that the VR user is currently interacting with using the controllers. \textit{Dynamic-Interaction} is a responsive implementation of the static condition, \textit{Interactive}, which enables the users to filter information depending on their hand motions.  

Based on a preference ranking and informal interviews, we decided to remove two visualisation techniques that participants liked the least for the main study: \textit{Head-Direction} and \textit{Dynamic-Interaction}. Regarding \textit{Head-Direction}, participants found that frequent changes to a VR user's line of sight and thus the visualisation made the technique confusing. Regarding \textit{Dynamic-Interaction}, participants found the visualisation difficult to understand as too little information was being shown. We also improved some visualisations, such as \textit{Dollhouse} which participants complained that the visualisation was too small to see clearly. We thus increased the scale of the visualisation to 1:5 ($\unit{1}{\meter}\times\unit{1}{\meter}$) from its original 1:10 ($\unit{0.5}{\meter}\times\unit{0.5}{\meter}$) scale.

\subsection{Selective Visualisation Framework}
Using our pilot study to refine our design, we selected four visualisation techniques for the main implementation of the study: 

\begin{itemize} 
    \item \textbf{Everything}: A VR-mimicking condition where the entire VE is visualised to the AR user at 1:1 scale. No modifications are made and the visualisation is symmetrical between the VR and AR user. 
    \item \textbf{Proximity}: An `arm's-reach' approach that dynamically visualises a \unit{1}{\meter} radius of VE around the VR user, with an additional \unit{0.5}{\meter} radius of decreasing opacity to fade-out the visible threshold of the technique. The parts of the VE visualised changes depending on the location of the VR user. 
    \item \textbf{Interactive}: A static, predetermined visualisation of interactive movable objects in the VE. This is a selection of objects that the VR user can pick up and interact with using their controllers. As a static visualisation, all the interactive objects are visualised at all times. 
    \item \textbf{Dollhouse}: A 1:5 scaled visualisation that provides a top-down overview of the VE, which hovers \unit{1}{\meter} above floor-level. The walls and ceiling of the virtual model are removed to facilitate looking into its interior.
\end{itemize}

\begin{figure} [h]
    \centering
    \includegraphics[width=\linewidth] {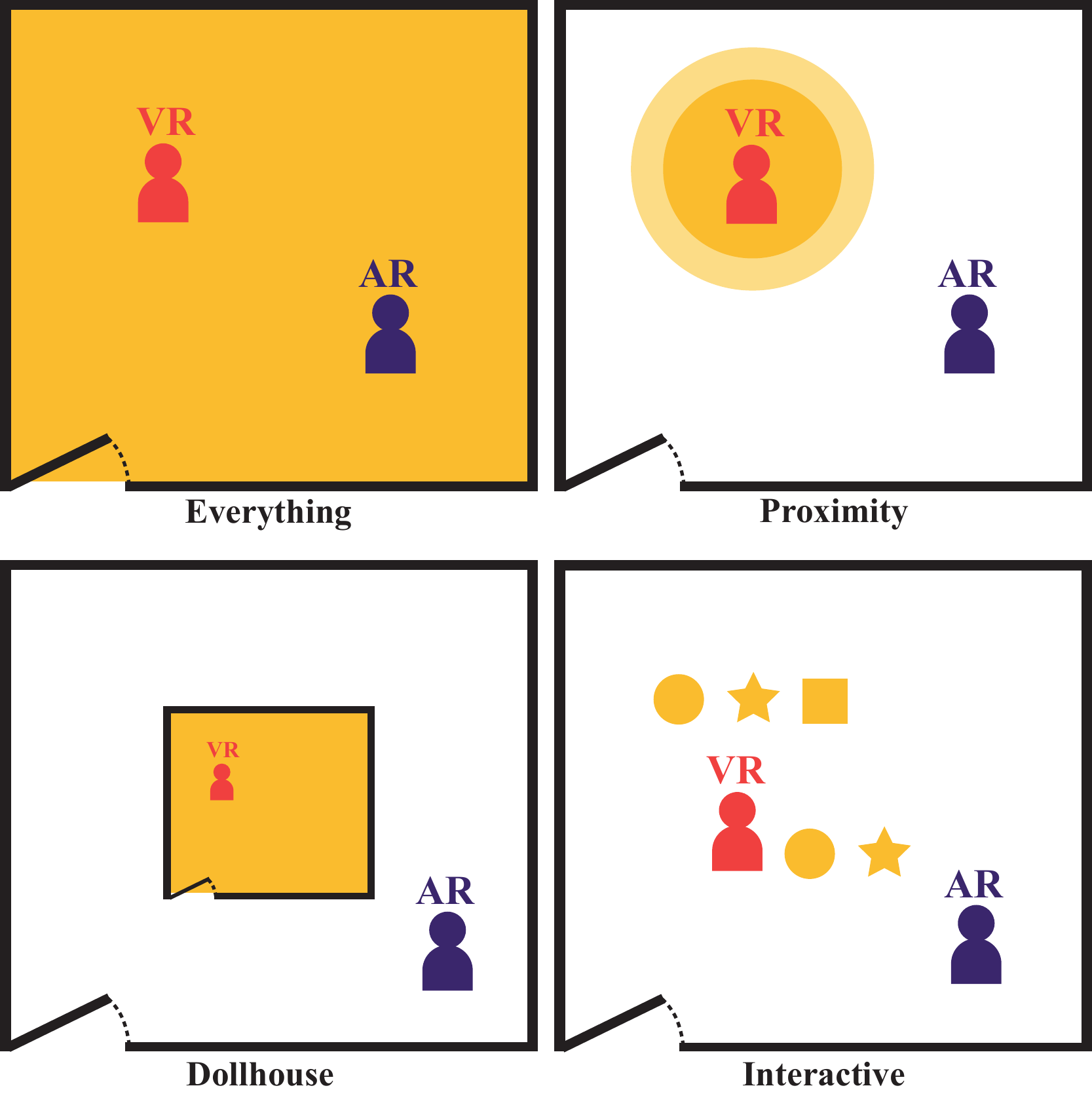}
    \caption{Selective Visualisation Techniques of VE: Entire VE is visible to the VR user in all conditions, while only coloured areas are visible to the AR user.}
    \label{fig:main_drawing}
\end{figure}

\begin{figure} [h]
    \centering
    \includegraphics[width=\linewidth] {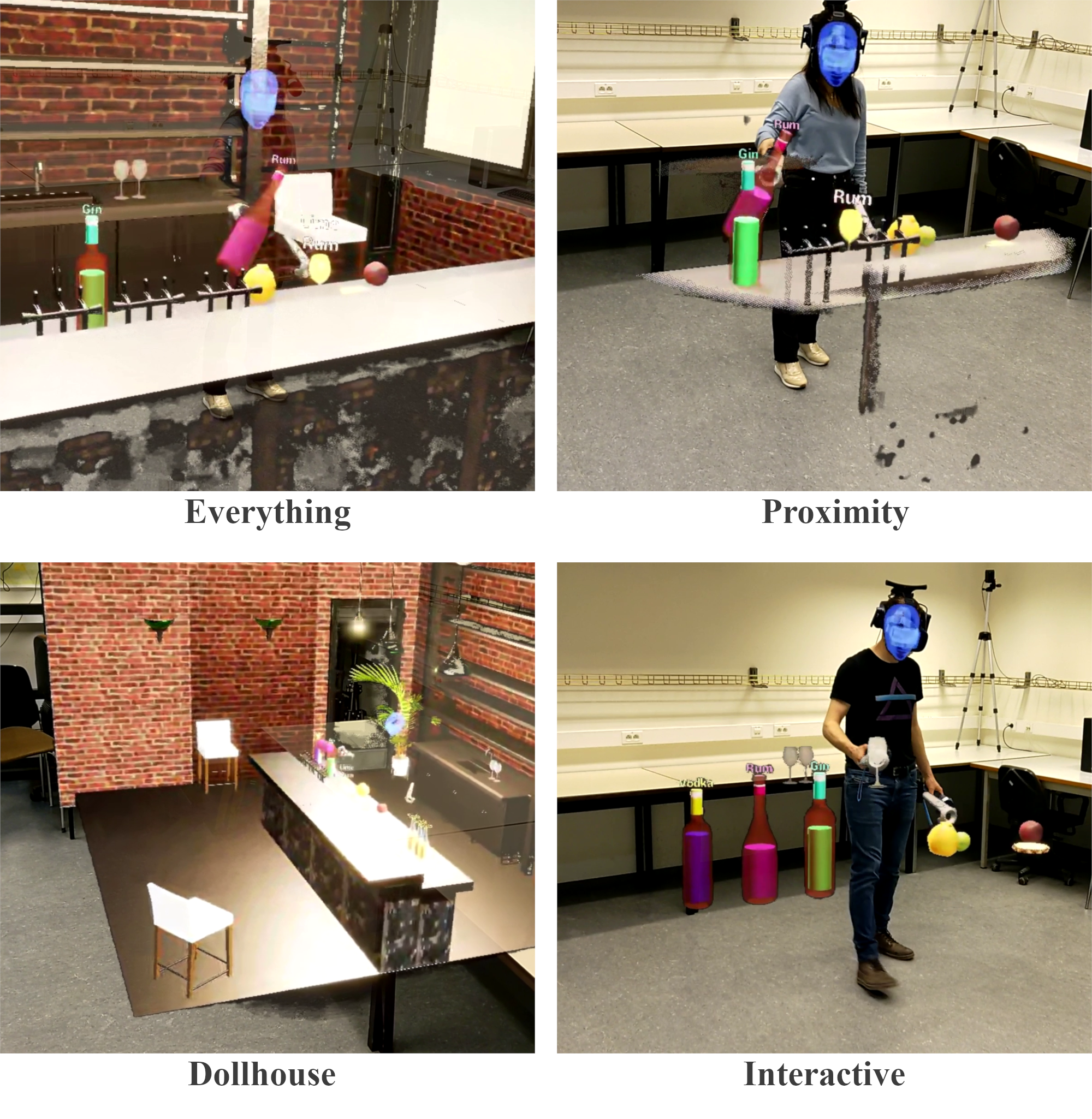}
    \caption{Selective Visualisation Techniques of VE (Mixed Reality Capture).}
    \label{fig:main_AR_perspective}
\end{figure}

These techniques can be seen as diagrams in \autoref{fig:main_drawing} and from the AR user's perspective in \autoref{fig:main_AR_perspective}.

\subsection{Implementation of System}
We developed this selective visualisation system in Unreal Engine 4.25 as a networked application, which runs on two computers on the same LAN with one as the server and the other as a client. We used a HTC Vive Pro and a Microsoft Hololens 2, both of which have their own coordinate systems: lighthouses managed by SteamVR and embedded camera-based tracking respectively. Using a custom calibration procedure the coordinate systems are aligned. This procedure consists of scanning two QR codes with the Hololens, and placing the HTC Vive controllers on top of these codes. With two corresponding points in both coordinate systems known, the origin of the Hololens coordinate system is transformed so that these points overlap with the corresponding points in the SteamVR coordinate system. In operation drifts between the coordinate systems can be observed up to a maximum of \unit{5}{\centi\meter}.

\subsection{The Virtual Environment}
For the purposes of testing the visualisation system, we created a `bartender simulation' as the VE. The participant assumes the role of the AR user within this simulation because we are investigating the AR user's perception of how a VR user interacts with a VE. The researcher assumes the role of VR user and conducts a `performance' using a predetermined script for the AR user to observe. This performance consists of making cocktails using three different recipes, and the order of the recipes and the actions performed for making them differed between the different visualisation techniques. The VR user used three types of interactive objects to perform this script: fruits, bottles and glasses. All these objects can be picked up and moved with the VR user's motion controllers. The glasses can hold slices of fruit and liquids. The contents of the glass are indicated by floating text above it. A fruit is added on entering the collision box of the glass. Liquids are only added when the top of the bottle collides with the glass, to mimic a pouring motion. When the glass is held upside-down the contents are emptied. On the bar counter there is floating text indicating the current order and a simplified three-ingredient recipe. Below this text is a collision box that checks the glass contents on collision, and when the contents are correct empties the glass and advances to the next recipe. These are the different events that the AR user can perceive in the VE that are triggered by the VR user. 

\section{Main User Study}

\subsection{Procedure}
\label{sec:study1_procedure}
We recruited thirteen participants for the main study, aged between 21 and 57 (M=30.62, SD=12.48; 6 male, 7 female). They had a low self-reported experience with VR and AR technologies (M=3.08, SD=1.19 on 7-point scale).

Participants were tasked with using the Hololens 2 to observe a VR user that is performing a bartender simulation. The Hololens was set to the highest brightness setting. \textit{Holographic remoting}\footnote{\url{https://docs.microsoft.com/en-us/windows/mixed-reality/develop/platform-capabilities-and-apis/holographic-remoting-player}} was used to stream the image from the computer to the Hololens. Participants were given an `event recognition task', a list of events which they need to recognize as they happen. These events are triggered by the VR user's actions. The researcher used a HTC Vive Pro with the Vive wireless attachment to perform the role of the VR user, following a predetermined set of actions on each trial. During each trial the VR bartender made three drinks, consisting of combining three ingredients in a glass each. Participants performed four trails, one for each technique, during which they could move around the lab to adjust their viewpoint. The study lasted about \unit{40}{\minute}.

Before taking part in the study, participants signed a consent form and filled in a demographics questionnaire, then we explained to them the event recognition task and instructed them on how to use the Microsoft Hololens 2. Before starting the first trial participants were given some time to look around the bar environment and get to know the positions of all the objects. The techniques were presented in counterbalanced order, using a balanced Latin square. During each trial participants were required to pay attention to the VR user and the VE. After each trial participants filled in which events they saw happen, Slater-Usoh-Steed's (SUS) presence questionnaire~\cite{Usoh2000UsingReality}, Kennedy's Simulator Sickness Questionnaire~\cite{Kennedy1993SimulatorSickness} and a questionnaire with custom questions on a 7-point likert scale. After the last trial participants were asked to rank the techniques (1\textsuperscript{st}, 2\textsuperscript{nd}, 3\textsuperscript{rd} and 4\textsuperscript{th}), and were interviewed on their thoughts on the techniques and the experience in general.

The study took place during the COVID-19 pandemic. The keyboard, mouse and desk area used by the participants were disinfected before and after the study, as well as the Hololens 2 for which a Cleanbox UV-C decontamination device\footnote{\url{https://cleanboxtech.com/}} was used. Participants and researcher disinfected their hands before and after the study, wore face masks and maintained a distance of at least \unit{1.5}{\meter} between them. There was at least \unit{30}{\minute} between participants to avoid them meeting and allow time to disinfect and ventilate our lab. The study and COVID-measures were approved by the university's privacy and ethics board (PRET).

\subsection{Results}
\begin{figure} [h]
    \centering
    \def\svgwidth{1\linewidth}
    \relscale{.8}
    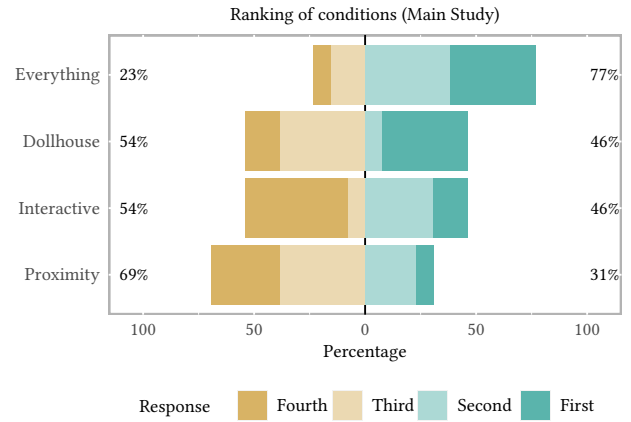
    \caption{Participant rankings of techniques in the main study.}
    \label{fig:pref1}
\end{figure}

\paragraph{\textbf{Preference Ranking}}
Participant preference of the techniques can be seen in \autoref{fig:pref1}. Following a pairwise Wilcoxon signed rank test, the \textit{Everything} technique was ranked significantly higher than \textit{Interactive} (p<0.05) and \textit{Proximity} (p<0.05), with 77\% ranking it at the first or second place. Both \textit{Interactive} and \textit{Dollhouse} were ranked first or second by 46\% of participants. Only 31\% of participants ranked \textit{Proximity} in first or second place.

\paragraph{\textbf{Event Recognition}}
We used a competence calculation (True Positive Rate - False Positive Rate)~\cite{Weller2005CulturalModel} to analyse how well the participants understood the VR user's actions during the event recognition task. `Competence' is the probability of knowing a correct answer without guessing and not by chance, and in this context refers to the probability of an AR user correctly identifying the actions conducted by the VR user. A Kruskal-Wallis test showed no significant difference (p=0.93) in competence across the four visualisation conditions. However, we observed a marginal difference in the mean competence that favoured the filtered visualisations. The competence values range up to 1, representing a participant that only indicated the correct events. The mean competence ranges between 0.65-0.76, and from highest to lowest: \textit{Interactive} (M=0.76, SD=0.32), \textit{Proximity} (M=0.71, SD=0.31), \textit{Everything} (M=0.67, SD=0.48), and \textit{Dollhouse} (M=0.65, SD=0.42).

\paragraph{\textbf{Interviews}} We analysed the interview using a thematic analysis~\cite{Braun2006UsingPsychology}. We categorised user responses in three themes: Firstly, the ability to focus on the simulation; secondly, the presence of the VR user; and thirdly, feedback on the visualisation methods. Participants found it harder to focus in the \textit{Everything} condition, with five participants finding real objects distracting and two finding virtual objects distracting. In contrast, five participants found it easier to focus in the more visually filtered \textit{Interactive} condition. This is higher than the number of people who stated that \textit{Dollhouse} or \textit{Proximity} helped focus, which was two. Regarding the presence of the VR user: five participants commented that rather than the physical appearance of the VR user, they found themselves focusing on the actions being conducted. Any mentions of the physical appearance of the VR user only arose from room-scale conditions, even if the VR user was co-located in all the conditions. Regarding feedback for the visualisation conditions: For \textit{Interactive}, six participants found they missed the bar counter as a point of reference in the scene. For \textit{Proximity}, two participants complained about having less control over the visibility of virtual artifacts, and two other participants about wanting to stay aware of the invisible part of the VE.

\paragraph{\textbf{SSQ and SUS}}
A Kruskal-Wallis test showed no significant difference (p=0.91) between the SSQ's Total Score, with the following mean values: control (M=9.81, SD=8.26), \textit{Dollhouse} (M=8.01, SD=9.96), \textit{Everything} (M=9.94, SD=10.6), \textit{Interactive} (M=6.64, SD=7.31) and \textit{Proximity} (M=7.57, SD=7.54). The SUS questionnaire was analysed by counting the number of 6 and 7 answers. There was no significant difference, though the mean for \textit{Everything} (M=1.15, SD=1.52) was higher than for the other conditions: \textit{Dollhouse} (M=0.08, SD=0.28), \textit{Interactive} (M=0.08, SD=0.28) and \textit{Proximity} (M=0.15, SD=0.38).

\section{Follow-up User Study}
\subsection{Changes to Selective Techniques}
Issues with the selective techniques were found in the main study, these were addressed and the resulting improved techniques evaluated in a follow-up study.

In feedback given during the interviews, participants mentioned issues with our selective techniques: for \textit{Interactive} they missed the bar counter as a point of reference and for \textit{Proximity} it was confusing that the environment disappeared completely which removed all context in which to see the highlighted area around the VR user. We thus implemented two new selective techniques to address this feedback: 
\begin{itemize}
    \item \textbf{Spotlight}: An improved version of \textit{Proximity} in which the VE in proximity of the VR user is rendered opaque, but modified to enable the AR user to see the rest of the VE as simple outlines. This allows users to see the highlighted area in context of the rest of the VE without it obstructing view of the physical environment. 

    \item \textbf{Context}: A refinement of \textit{Interactive} that responds to the participants' desire to see more of the VE. The furniture that supports the interactive objects can now be seen, i.e.\ the counter and sink. 
\end{itemize}

\begin{figure} [h]
    \centering
    \includegraphics[width=\linewidth]{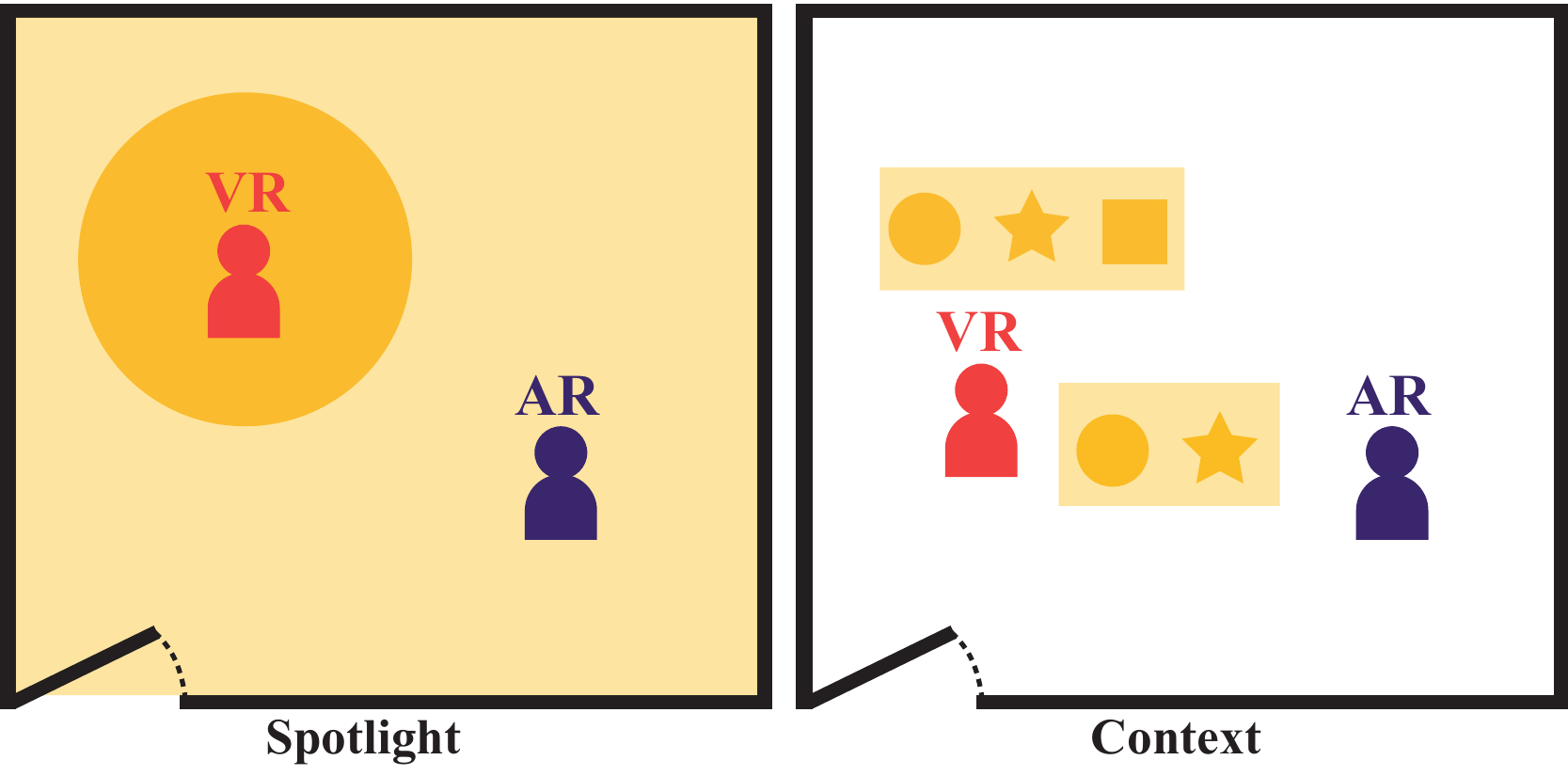}
    \includegraphics[width=\linewidth] {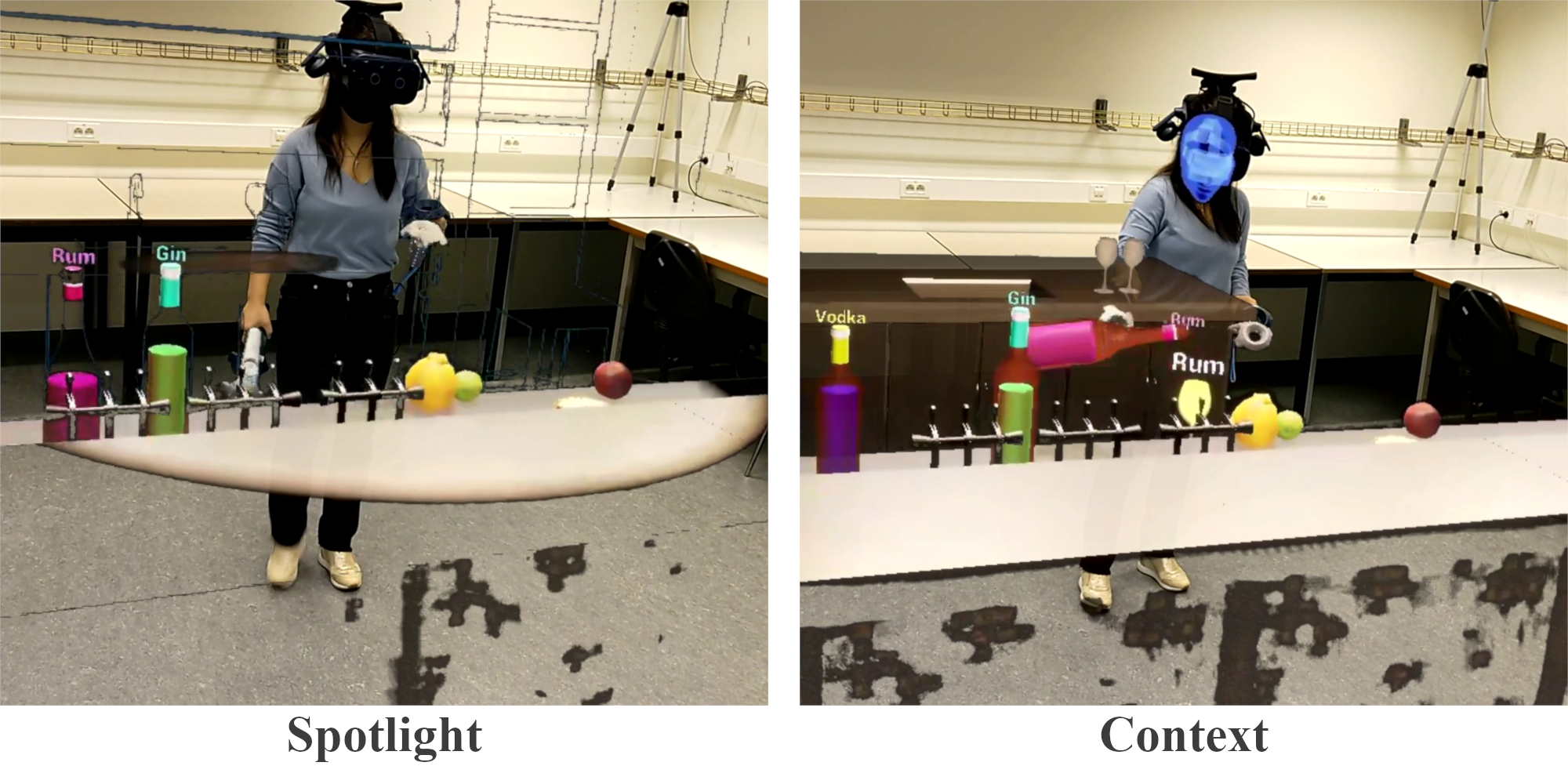}
    \caption{Improved Selective Visualisation \textit{Spotlight} (left) and \textit{Context} (right) Techniques diagrams (top) and as Mixed Reality Capture (bottom). The whole VE is visible to the VR user in all conditions, while only coloured areas are visible to the AR user (diagram).}
    \label{fig:follow_combined}
\end{figure}

These techniques can be seen as diagrams and from the AR user's perspective in \autoref{fig:follow_combined}.

\subsection{Procedure}
The follow-up study followed the same procedure as the first study described in \autoref{sec:study1_procedure}, with the \textit{Proximity} technique replaced by \textit{Spotlight} and the \textit{Interactive} technique replaced by \textit{Context}.

For the follow-up study we recruited 13 participants, aged between 19 and 57 (M=29.77, SD=14.35; 6 male, 7 female). With low self-reported experience with VR and AR (M=2.92, SD=1.25 on a 7-point scale).

\subsection{Results}
\begin{figure} [h]
    \centering
    \def\svgwidth{1\linewidth}
    \relscale{.8}
    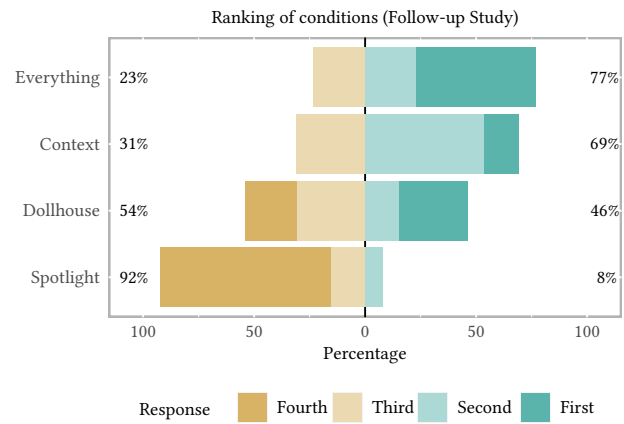
    \caption{Participant rankings of techniques in the follow-up study.}
    \label{fig:pref2}
\end{figure}

\paragraph{\textbf{Preference Ranking}}
Participant preferences can be seen in \autoref{fig:pref2}. A pairwise Wilcoxon signed-rank test showed that \textit{Spotlight} ranked significantly lower than the other techniques (p<0.01 for \textit{Everything} and \textit{Context}, and p<0.05 for \textit{Dollhouse}) with 92\% of participants ranking it third or fourth. \textit{Dollhouse} was ranked significantly lower (p<0.05) than \textit{Everything} with 54\% of participants ranking it third or fourth, and 77\% ranking \textit{Everything} first or second. 69\% of participants ranked \textit{Context} first or second.

\paragraph{\textbf{Event Recognition}}
There were no significant differences in how well participants could recognise events (Kruskal-Wallis test,  p=0.72). However competence values for selective conditions were marginally higher. The mean competence ranges 0.67-0.83, and from highest to lowest:  \textit{Context} (M=0.83, SD=0.25), \textit{Everything} (M=0.76, SD=0.43), \textit{Spotlight} (M=0.75, SD=0.31) and \textit{Dollhouse} (M=0.67, SD=0.43).

\paragraph{\textbf{Interviews}}
The same three themes were identified in the interviews as in the main study: ability to focus on the task, the presence of the VR user and feedback on the methods. Four participants found the VE distracting in \textit{Everything}, while three participants were distracted by the VE represented as outlines in \textit{Spotlight} finding that the out-of-focus environment was unclear. Three participants also found that \textit{Spotlight} helped them focus more on the bartender. Four participants only saw the bartender as an avatar, while six others mentioned that they could see the physical person behind the avatar in the room-scale conditions. Three participants said they could not see the bartender well enough. Six participants found \textit{Everything} and \textit{Context} very similar, five participants even expressed difficulty in discerning these two techniques. Three participants expressed frustration with the limited vision in \textit{Spotlight}. Nine participants found that \textit{Dollhouse} gave them a good overview of the VE, five participants found it too small.

\paragraph{\textbf{SSQ and SUS}}
A Kruskal-Wallis test showed no significant difference (p=0.96) between Total Score on Simulator Sickness, with the following mean values: control (M=4.72, SD=13.65), \textit{Dollhouse} (M=3.21, SD=7.67), \textit{Everything} (M=2.75, SD=5.17), \textit{Context} (M=1.86, SD=3.47) and \textit{Spotlight} (M=4.09, SD=10.19). The SUS questionnaire was analysed by counting the number of 6 and 7 answers. There was no significant difference (p=0.52), though the mean for \textit{Everything} (M=1.46, SD=2.22) was higher than for the other conditions: \textit{Dollhouse} (M=0.62, SD=1.45), \textit{Context} (M=0.77, SD=1.69) and \textit{Spotlight} (M=0.62, SD=1.19).

\section{Discussion}
We hypothesised that it would be possible to remove parts of the VE that are non-essential to the task being performed in it without altering an external user's perception of the task itself. Our results indicated that removing a large part of the VE indeed does not create a significant difference in how well an AR user can identify the events triggered by a VR user. However, our findings also reveal that competence does not necessarily correspond with user preference on the different visualisations, and we identified that participants preferred to see the supporting furniture of visible objects and did not prefer to lose control over the visualisation.

An initial static selection \textit{Interactive} was also not preferred. Participants indicated that they could not see enough of the VE. We developed an improved iteration, \textit{Context}, which showed relevant furniture in addition to the original objects of the VE. Participants expressed a more positive response for this visualisation, commenting that \textit{Context} was similar to seeing the entire VE. Some of the participants even expressed being unable to tell the difference between this selection and seeing \textit{Everything}. Between the main and follow-up study, the preference ranking of this static visualisation has increased by one rank. 

On the effect of scale we found that \textit{Dollhouse} provided a better overview of the VE, but also that many participants found it too small. We were able to use a 1:5 scale because our VE was only as large as the physical size of the room. Larger VEs need to be scaled down more, which can make the issue of them being too small worse, or not shown entirely which can make users lose their overview on the VE. In ObserVAR~\cite{Thanyadit2019ObserVAR:Reality} participants found the \textit{World Scale} condition easier to use than the \textit{World in Miniature} condition which is supported by our results where participants preferred the \textit{Everything} and static selection room-scale techniques over the \textit{Dollhouse} technique. The results from ObserVAR indicate that their \textit{World Scale} provided a better overview, which contradicts our results that indicated \textit{Dollhouse} as the technique providing a better overview. This can be explained by the ObserVAR implementation of \textit{World in Miniature} that visualises a separate miniature for each VR user, thus splitting up the information required by their user study participants.

Two types of selection were investigated, a predetermined static selection of objects and a dynamic selection that follows the VR user. Participants did not prefer the dynamic selection, citing lack of control over what they could see in the VE. A similar trend was cited in HMD light, in which external users looking into a VR user's VE wanted to have more control over the visualisation of the VE \cite{Wang2020HMDInteraction}. Further comparisons can be made with HMD Light regarding user preferences on the stability of the visualisation. Comparing a third person view with a first person view, most users in HMD Light chose third person view because as it was more stable and holistic than the 1st person view \cite{Wang2020HMDInteraction}. In our study, more holistic and static visualisations such as \textit{Everything} and \textit{Dollhouse} were preferred over more dynamic visualisations such as \textit{Proximity} or \textit{Spotlight}. 

Participants in selective visualisations such as \textit{Interactive} or \textit{Context} could identify VE events more accurately than seeing \textit{Everything}. Compared to other visualisations, the majority of the participants highlighted some distracting features in the \textit{Everything} visualisation. In contrast, a number of participants cited \textit{Interactive} in particular as useful for maintaining focus, despite the lower preference rating compared to \textit{Everything}. However, it is important to note that statistically we have found no significant differences proving that more selective visualisations improve focus. We have only observed that the mean values for competence are marginally higher for the selective visualisations in this instance of a `bartender' VE. Further investigation is necessary, perhaps with a range of different levels of information that incorporate tasks of greater complexity.

For researchers and developers in CR, we recommend different visualisations depending on the purpose and appearance of the VE. For a VE that requires the AR user to have an overview of the space, the \textit{Dollhouse} condition has shown to be the most effective of those evaluated in this study. It is important to note that the scale of the \textit{Dollhouse} depends on the size of the VE, as very large VEs are potentially limited by the physical space available even when scaled, and there exists a limit to how small a VE can be visualised before the AR user no longer understands what the VE represents. Additionally, the VE should be visualised as a static selection as opposed to a dynamic selection whenever possible, as static selections have shown to rank higher in terms of user preference. Lastly, it is possible to remove all non-essential information and preserve the recognition of events, but showing the immediate context matters for user preference. 

\section{Conclusion and Future Work}
In two studies we investigated how a selective visualisation system of VEs can influence an AR user's perception of a co-located VR user. We looked at two variables: the level of visual information and the effect of scale. Regarding level of visual information, we observed that filtering specific selections of the VE did not significantly affect the competence of how well people could identify events in the VE. These selections were based on the interactive range and possibilities of the VR user. Regarding scale, users generally agreed that smaller visualisations provide a better overview of the VE, but had the chance of decoupling the user from the task at hand. In terms of user preference, our qualitative data showed that participants tended to prefer static visualisations over dynamic visualisations, disliking the lack of control they could exercise for visualising the VE. 

In future work we would like to improve these visualisations to apply to a greater variety of VE contexts. Techniques such as \textit{Proximity} are generalisable, but techniques such as \textit{Context} are very specific to the context of the VE as they use a predetermined selection of virtual objects. This selection of visualised objects can be made in different ways, instead of a predetermined selection future work can investigate the creation of an interface for the AR, or VR, user to make this selection themselves.

Moreover, we would like to apply these visualisations into an interactive implementation of this visualisation system, as currently the AR user only assumes a passive spectator role in the task. We could test our visualisations in a collaborative task that requires both the AR user and VR user to interact with elements from the VE.

\begin{acks}
This research is supported by \textit{Internal Funds KU Leuven} (C14/20/078).
\end{acks}

\bibliographystyle{ACM-Reference-Format}
\bibliography{main}

\end{document}